\documentclass[a4paper,12pt]{article}

\newcommand{\cv}[1]{{#1}}

\usepackage[T2A]{fontenc}
\usepackage{amsmath,amssymb}
\usepackage{cite}
\usepackage[linktocpage=true,plainpages=false,pdfpagelabels=false]{hyperref}

\usepackage[utf8]{inputenc}
\usepackage[english]{babel}

\usepackage{pgfplots}

\newcommand{\dd}{\partial}
\newcommand{\de}{\delta}
\newcommand{\De}{\Delta}
\newcommand{\m}{\mu}
\newcommand{\n}{\nu}
\newcommand{\ls}{\left(}
\newcommand{\rs}{\right)}
\newcommand{\al}{\alpha}
\newcommand{\ff}{\varphi}
\newcommand{\ta}{\tau}
\newcommand{\te}{\theta}
\newcommand{\ti}{\tilde}
\newcommand{\om}{\omega}
\newcommand{\si}{\sigma}
\newcommand{\dz}{\zeta}
\newcommand{\str}[1]{\mathrel{\mathop{\longrightarrow}\limits_{#1}}}

\newcommand{\disn}[2]{$$\displaylines{\refstepcounter{equation}%
            \label{#1}\hskip 1em minus 1em #2\hfilneg}$$}
\newcommand{\nom}{\hfil\hskip 1em minus 1em (\theequation)}
\newcommand{\no}{\hfil \hskip 1em minus 1em\phantom{(\theequation)}%
            \hfilneg\cr\hfilneg\hskip 1em minus 1em\hfil}
\newcommand{\ns}{\hfill\cr\hfill}

\begin{document}

\title{Searching for the classical \cv{version} of Hawking radiation\\
and screening of Coulomb field by the horizon}

\author{
S.~A.~Paston\thanks{E-mail: pastonsergey@gmail.com},
D.~S.~Shatkov\thanks{E-mail: ya.danilishe@yandex.ru}\\
{\it Saint Petersburg State University, Saint Petersburg, Russia}
}
\date{\vskip 15mm}
\maketitle

\begin{abstract}
We investigate the possibility of the existence of a classical \cv{version} of Hawking radiation -- solutions to classical field equations that consist solely of outgoing waves, in the spacetime of a collapsing black hole. The non-static nature of the corresponding metric results in the absence of energy conservation for matter, which could otherwise a priori prohibit such solutions. A specific and simple scenario is considered: a black hole formation as a result of the collapse of a thin shell, which is not necessarily dust-like. In the corresponding spacetime, we study solutions of the equations for a real massless scalar field that take the form of purely outgoing waves. In addition to the homogeneous equation, we also examine the case of a constant point source of the field located at the symmetry center. The general solution outside the shell is expressed in terms of the confluent Heun function, while the equations inside the shell and the matching conditions at its surface are formulated as an integral equation. The analysis of various solution asymptotics enables the reduction of the integral equation to a matrix equation, which is subsequently solved numerically.
\end{abstract}

\newpage
\section{Introduction}
Hawking radiation, first predicted in the pioneering work \cite{hawking-1974}, despite its low probability of direct experimental detection, is a critically important theoretical concept in black hole (BH) physics. This significance arises largely from the thermodynamic properties of BHs (see, e.g., \cite{ongbhtherm, solodbhentr}) and the frequently discussed information paradox (see, e.g., \cite{rajuinform}). An important implication of Hawking radiation is the idea of BH evaporation, according to which the mass of a BH decreases over time. Determining how a BH's mass changes with time is a complex problem (see, e.g., \cite{goodevapor} and references therein), as it requires accounting for \textit{backreaction} -- the influence of radiation on the spacetime metric.

Hawking radiation is understood to be an inherently quantum process. At first glance, it seems entirely evident that within the framework of classical (non-quantum) physics, no radiation can occur, as "nothing can escape from a black hole". However, it is important to consider this more carefully. Specifically, we must distinguish between two scenarios: a black hole formed via collapse and an eternal black hole, such as a manifold with the metric of the maximal analytic extension of the Schwarzschild solution, entirely described in Kruskal-Szekeres coordinates \cite{kruskal, szeksingul}.

\begin{figure}[htbp]\centering
\includegraphics[height=0.3\textwidth]{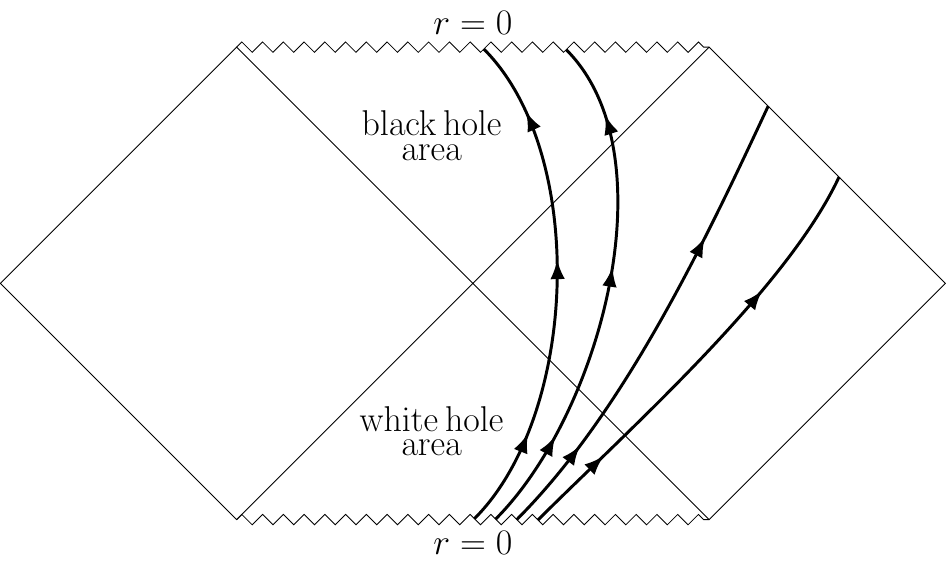}
\caption{Penrose diagram for an eternal spherically symmetric uncharged black hole.
The arrows indicate the possible flow of radiation emerging from the "white-hole" singularity.}
\label{penrvech}
\end{figure}

In the case of an eternal black hole, the manifold contains a region corresponding to a "white" hole (the Penrose diagram for such a BH is shown in Fig.~\ref{penrvech}).
This region is separated from the regular spacetime by a horizon, through which particles and field perturbations can pass from below the horizon outward, but not vice versa. This region contains a central "white-hole" singularity at $r=0$, which could, in principle, serve as a source of any type of particles or radiation, with its parameters determined by the choice of initial conditions at the singularity. Therefore, in the framework of classical physics, an external observer could detect radiation from an eternal black hole with relatively arbitrary properties (see Fig.~\ref{penrvech}).

\begin{figure}[htbp]\centering
\includegraphics[height=0.3\textwidth]{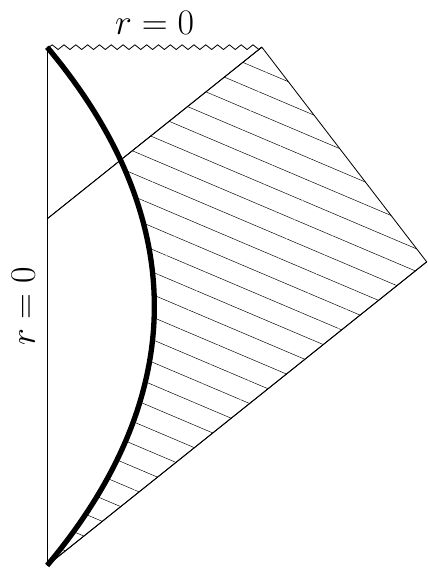}
\caption{Penrose diagram for a spherically symmetric uncharged black hole formed via collapse.
The vertical line corresponds to the origin of coordinates $r=0$.}
\label{penr}
\end{figure}

More specific results can be obtained for the case where a black hole forms as a result of collapse. In this scenario, there is no "white-hole" region and thus no "white-hole" singularity. Nevertheless, the argument "nothing can escape from a black hole" is insufficient, as the black hole did not exist in the past, and an external observer can detect radiation originating from this past. The Penrose diagram for a BH formed via collapse is shown in Fig.~\ref{penr}.
It is worth noting that in the case of collapse, for an external observer, at any finite moment in time (according to their clock), the black hole \emph{has not yet formed} in the sense that the signals they receive from the collapsing matter are always emitted \textit{before} the matter crosses its Schwarzschild radius.

The reasoning for why, in the absence of collapsing matter, no radiation would occur is straightforward -- it follows from the law of energy conservation. In Minkowski spacetime, if we seek solutions to equations describing some physical radiation, such as Maxwell's equations, energy conservation ensures that if there was no radiation in the distant past, there will be none in the distant future. Thus, the key role is played by energy conservation. However, in the presence of a varying gravitational field, the energy of matter alone is not conserved -- matter can exchange energy with the gravitational field. To ensure energy conservation for matter, the manifold must possess a timelike Killing vector (see, e.g., \cite{wald}), which effectively means that the metric is time-independent in some coordinate system. For the manifold shown in Fig.~\ref{penr}, corresponding to collapse, this condition is not satisfied -- the metric is fundamentally non-static. Therefore, there is no \textit{a priori} reason to believe that field equations on such a manifold cannot yield solutions where radiation is absent in the distant past but present in the future.

It is important to distinguish between two cases: when such radiation occurs in the future for a limited time and when it persists indefinitely. The latter case can be considered a classical \cv{version} of Hawking radiation, as ordinary Hawking radiation does not dissipate over time. In the first case, the resulting radiation can be interpreted as a consequence of the changing metric during BH formation, after which the system transitions into a "steady-state" condition from the perspective of an external observer.
\cv{In the second case, the cause of the radiation is solely the geometry of spacetime and the neglect of backreaction. However, if backreaction were to be taken into account, energy would start to be conserved, and we could observe the impact of the classical radiation process on the black hole's geometry, which would allow for a discussion of its evaporation process.
}

In this work, we attempt to find a solution where radiation is absent in the distant past but present in the future. To specify the problem, we need to select field equations describing the radiation and a collapse regime. For simplicity, we consider a real, massless, free scalar field $\varphi$. In the absence of sources, the field's equation of motion is the d'Alembert equation on a manifold with a given metric $g_{\mu\nu}$, determined by the collapse regime:
 \disn{vv1}{
g^{\mu\nu}D_\mu\partial_\nu \varphi=0,
\nom}
where $D_\mu$ is the covariant derivative. For simplicity, we assume the collapse of a spherically symmetric thin shell and restrict our analysis to spherically symmetric solutions (i.e., those corresponding to the zeroth spherical harmonic). We do not assume that the shell consists specifically of dust-like matter, allowing for arbitrary transverse tension in the shell material. This flexibility allows us to conveniently choose the shell's collapse regime.

As will be shown below, solving the problem becomes simpler if we generalize it to include a constant point source for the field $\varphi$ at the center of symmetry. This means we seek a solution not of \eqref{vv1} but of the more general equation:
 \disn{vv2}{
g^{\mu\nu}D_\mu\partial_\nu \varphi=4\pi Q \delta(x^i),
\nom}
which reduces to \eqref{vv1} when $Q=0$. Inside the shell (where the metric corresponds to flat spacetime), the transition from \eqref{vv1} to \eqref{vv2} corresponds to adding a Coulomb field $\varphi=Q/r$ to $\varphi$. Thus, solving the problem for $Q\neq0$ allows us to address the new question: how does horizon formation affect the behavior of the Coulomb field as observed from afar?

In Section~\ref{razd-kl}, we describe the metric corresponding to the collapse of a thin spherical shell and fix its collapse regime in a convenient way. Section~\ref{razd-kr} addresses the field equations inside the shell and derives the conditions for field behavior at the shell. In Section~\ref{razd-sn}, the field equations outside the shell are analyzed, and an integral equation is constructed to determine the desired solution that corresponds to outgoing waves at large distances from the BH, with no incoming waves. The solution is expressed in terms of the confluent Heun function. To enable accurate numerical solutions of the integral equation, various asymptotics of the field equations and the kernel of the integral equation are derived in Section~\ref{razd-as}. The integral equation is then reduced to a matrix equation in Section~\ref{razd-ob}. Numerical results are presented in Section~\ref{razd-rz} and discussed in the concluding Section~\ref{razd-za}.

\section{Description of the collapse of a thin spherical shell}\label{razd-kl}
Let us describe the metric corresponding to the collapse of a thin spherically symmetric shell.
A similar problem was considered, for instance, in \cite{akhmhawk}.
Outside the shell, the metric is well-known; the corresponding interval in Schwarzschild coordinates is given by
 \disn{kl1}{
ds^2=\ls 1-\frac{R}{r}\rs dt^2-\frac{dr^2}{1-\frac{R}{r}}-r^2 \ls d\te^2+\sin^2\te d\varphi^2\rs,
\nom}
where $R$ is the Schwarzschild radius related to the mass of the shell (we assume the speed of light $c=1$).
Inside the shell, the metric should have the same form but with $R=0$ since there is no mass under the shell,
i.e., it should be expressed as the Minkowski metric in spherical coordinates:
 \disn{kl2}{
ds'^2=dt'^2-dr'^2-r'^2  \ls d\te^2+\sin^2\te d\varphi^2\rs.
\nom}
The spherical coordinates $t',r'$ used here are marked with primes to emphasize that they do not continuously match the Schwarzschild coordinates $t,r$ used in \eqref{kl1},
as the metrics \eqref{kl1}, \eqref{kl2} obviously do not coincide on the shell.
Here and further, when discussing different coordinate systems, we omit the angular coordinates $\te, \varphi$ for brevity since they are always identical.

\cv{
The shell is a hypersurface that divides the entire spacetime into regions both inside and outside of it. Therefore, it is necessary to write down some junction conditions for it \cite{mars_1993}.}
The metric must be continuous on the shell when using some internal coordinates $t,r$ that continuously match the Schwarzschild coordinates outside the shell.
Thus, there exists a certain diffeomorphism within the shell:
 \disn{kl3}{
t,r\quad\longrightarrow\quad t'(t,r),r'(t,r).
\nom}
This diffeomorphism allows us to find the components of the metric $g_{\m\n}$ inside the shell in $t,r$ coordinates:
 \disn{kl4}{
g_{00}=\ls\frac{\dd t'}{\dd t}\rs^2-\ls\frac{\dd r'}{\dd t}\rs^2,\qquad
g_{11}=\ls\frac{\dd t'}{\dd r}\rs^2-\ls\frac{\dd r'}{\dd r}\rs^2,\qquad
g_{01}=\frac{\dd t'}{\dd t}\frac{\dd t'}{\dd r}-\frac{\dd r'}{\dd t}\frac{\dd r'}{\dd r},\no
g_{22}=-r'^2,\qquad
g_{33}=-r'^2 \sin^2\te,
\nom}
as it is known that in $t',r'$ coordinates the metric $g'_{\m\n}$ corresponds to the interval \eqref{kl2}.
The requirement that the metric is continuous on the shell in $t,r$ coordinates leads to the following relations on the shell:
$r'=r$ (which is geometrically evident) and
 \disn{kl5}{
\ls\frac{\dd t'}{\dd t}\rs^2-\ls\frac{\dd r'}{\dd t}\rs^2=1-\frac{R}{r},\qquad
\ls\frac{\dd t'}{\dd r}\rs^2-\ls\frac{\dd r'}{\dd r}\rs^2=-\frac{1}{1-\frac{R}{r}},\qquad
\frac{\dd t'}{\dd t}\frac{\dd t'}{\dd r}-\frac{\dd r'}{\dd t}\frac{\dd r'}{\dd r}=0.
\nom}
It should be noted that a point on the shell corresponds to a certain worldline, and this condition must hold at all points of this line. Assuming the shell is always contracting,
it is convenient to parametrize the points of this worldline by the value of $r$. The collapsing regime of the shell can then be described by the dependence of $t$ on $r$. Denote this dependence as $\bar t(r)$ (or $\bar t'(r')$ in primed coordinates).

The requirement that $r'$ and $r$ coincide at all points of the shell's worldline (we always imply a point on the shell with fixed angular coordinates $\te$ and $\varphi$) can be expressed as the condition:
 \disn{kl6}{
r'(\bar t(r),r)=r
\nom}
\cv{(here, $\bar t(r)$ corresponds to the worldline of the shell),}
which must hold for any $r$. Differentiating this with respect to $r$, we obtain:
 \disn{kl7}{
\frac{\dd r'}{\dd t}\frac{1}{v(r)}+\frac{\dd r'}{\dd r}=1,
\nom}
where the quantity $v(r)=(d\bar t(r)/dr)^{-1}$ represents the velocity of the shell. Since we assume the shell is always contracting, it follows that $v(r)<0$.

It can be observed that the four equations \eqref{kl5}, \eqref{kl7} allow us to determine the values of all partial derivatives involved at the shell. Since these equations are quadratic, there is ambiguity in choosing the signs, which must be resolved by requiring $\dd t'/\dd t>0$, $\dd r'/\dd r>0$.
The resulting expressions are straightforward but cumbersome; they are presented in Appendix~1.

In addition to the continuity of the metric on the shell, the Israel junction condition \cite{israelshell} must also hold, constraining the normal derivatives of the metric.
This condition involves the surface energy-momentum tensor (EMT) density of the matter composing the shell. Since no specific assumptions are made about the properties of this matter, there are no restrictions on the normal derivatives of the metric. Consequently, the collapse velocity function $v(r)$ can be arbitrarily specified (it is only necessary to ensure that the worldline of the shell is timelike). Then, the Israel condition uniquely determines the EMT of the shell. If specific physical properties of the shell material (e.g., dust-like matter) were assumed, this would impose constraints on the EMT, leading to an equation for $v(r)$, meaning the shell's properties determine its motion.
\cv{We assume that the properties of the matter forming the shell (the law governing its transverse pressure) are not relevant for the problem under consideration. Therefore, we do not impose any restrictions on the transverse pressure, but instead, we choose the collapse velocity function $v(r)$.
}

The worldline of the shell can be determined for a given function $v(r)$ (recall that $v(r)<0$):
 \disn{kl8}{
\bar t(r)=\int \frac{dr}{v(r)}.
\nom}
The condition of its timelike nature for $r>R$, i.e., outside the horizon, is given by
 \disn{kl9}{
\ls 1-\frac{R}{r}\rs -\frac{v(r)^2}{1-\frac{R}{r}}>0\quad\Rightarrow\quad |v(r)|<1-\frac{R}{r}.
\nom}
To further specify the collapse conditions, we assume that the collapsing thin shell reaches its Schwarzschild radius in finite Eddington-Finkelstein time $\ta=t+r+R\log((r-R)/R)$.
This imposes the condition $v(r)/(r-R)\rightarrow -1/R$ as $r\to R$ on the behavior of $v(r)$.

For simplicity, let us choose the collapse-defining function $v(r)$ in a straightforward form:
 \disn{kl10}{
v(r)=-\frac{1}{2} \ls 1-\frac{R^2}{r^2}\rs.
\nom}
It is easy to verify that this choice satisfies both the timelike condition \eqref{kl9} and the finite Eddington-Finkelstein time condition.
The corresponding expression for the shell's worldline, according to \eqref{kl8}, is:
 \disn{kl11}{
\bar t(r)=-2r-R \log\frac{2(r-R)}{e(r+R)}.
\nom}
Here, $e$ is the base of the natural logarithm, and the additive constant is chosen for convenience in further calculations.
\cv{The choice of the formula \eqref{kl10} corresponds to the infinite radius of the shell as \(t \to -\infty\), i.e., the fall of the shell from infinity. The question of how the results would change with a different choice of the shell's collapse mode, for example, when it has a finite radius as \(t \to -\infty\), requires a further study.
}

The Penrose diagram for the manifold described in this section is shown in Fig.~\ref{penr}.
The shell's worldline is depicted by a bold line. The region to its left has the flat metric \eqref{kl2}, while the region to its right has the Schwarzschild metric \eqref{kl1}.

\section{Solution inside the shell and constraints on it}\label{razd-kr}
We will study wave propagation corresponding to a free real massless scalar field $\ff$ in the spacetime described in the previous section. We will focus on the simplest case of spherically symmetric waves.

Recall that the d'Alembert equation \eqref{vv1} can be rewritten as
\disn{kr1}{
\dd_\m\ls\sqrt{-g}\,g^{\m\n}\dd_\n\ff\rs=0.
\nom}
This second-order equation can be solved separately inside and outside the shell, but it is necessary to impose matching conditions on the shell -- continuity of the field $\ff$ along with its first derivatives. Naturally, this continuity must be ensured in the coordinates in which the metric is continuous, i.e., the coordinates $t, r$ described in the previous section. Since we are interested in solutions that describe waves propagating outside the horizon and observable at large values of $r$ (such solutions can be considered as classical \cv{version} of Hawking radiation), we will look for solutions only in the region $r > R$.

Consider the equation \eqref{kr1} inside the shell in the coordinates $t', r'$, where the metric corresponds to \eqref{kl2}.
\cv{Any solution to the wave equation in Minkowski space can be written as a superposition of plane waves. However, as mentioned above, we will only be interested in spherical waves.
For the field $\ff'(x')$, under the assumption of spherical symmetry, the equation \eqref{kr1}} takes the form
\disn{kr2}{
\dd^2_{t'}\ff'-\frac{1}{r'^2}\dd_{r'}\ls r'^2\dd_{r'}\ff'\rs=0.
\nom}
Note that the left-hand side of this equation represents the action of the d'Alembert operator $\square=\dd^2_{t'}-\Delta$ on the field $\ff'$, expressed in spherical coordinates. By introducing a new field $\ti\ff'(x') \equiv r'\ff'(x')$ and performing straightforward calculations, this equation can be reduced to the simpler form
\disn{kr3}{
\dd^2_{t'}\ti\ff'-\dd^2_{r'}\ti\ff'=0
\nom}
of the two-dimensional wave equation.

Its solution has the standard form of a combination of two traveling waves:
\disn{kr3.1}{
\ti\ff'(t',r')=a(t'+r')+b(t'-r'),
\nom}
where $a$ and $b$ are arbitrary sufficiently smooth functions. Since $\ff' = \ti\ff'/r'$, it follows that if $\ti\ff'|_{r'=0} \ne 0$, then $\ff'$ exhibits Coulomb-like asymptotics at the origin. In this case, $\Delta\ff'$ would have a term proportional to $\de(x^i)$, and the equation \eqref{kr2} (and hence \eqref{kr1}) would not be satisfied. Therefore, to satisfy the equation \eqref{kr2}, it is additionally required that the condition $\ti\ff'(t',r')|_{r'=0}=0$ holds, leading to $b = -a$, i.e.,
\disn{kr4}{
\ti\ff'(t',r')=a(t'+r')-a(t'-r').
\nom}

Such a form of the solution inside the shell imposes certain constraints on the behavior of the values of the field $\ti\ff'$ and its derivatives at the shell itself.
\cv{We will limit ourselves to finding these constraints, but will not explicitly recover the solution inside the shell, since this is not necessary for the analysis of solutions outside the shell.}
It turns out to be simplest to first derive such constraints, which are necessary but, generally speaking, not sufficient conditions for expressing the solution of equation \eqref{kr3} in the form of \eqref{kr4} inside the shell. First, note that from the behavior \eqref{kr4}, it follows that
\disn{kr5}{
\dd_{t'}\ti\ff'(t',r')+\dd_{r'}\ti\ff'(t',r')=2\dot a(t'+r'),\no
\dd_{t'}\ti\ff'(t',r')-\dd_{r'}\ti\ff'(t',r')=-2\dot a(t'-r'),
\nom}
where the dot denotes differentiation of the function with respect to its argument. In particular, these equations hold on the shell, i.e., at $t'=\bar t'(r')$ (see after \eqref{kl5}). If we introduce a function $\hat r(r)$ such that for all $r$, the following relation holds:
\disn{kr6}{
\bar t'(\hat r(r))+\hat r(r)=\bar t'(r)-r,
\nom}
(henceforth, we take into account that on the shell $r'=r$), then from \eqref{kr5}, the function $a$ can be eliminated, resulting in a relation that holds at all points of the world line of the shell:
\disn{kr7}{
\ls\dd_{t'}\ti\ff'(t',r')+\dd_{r'}\ti\ff'(t',r')\rs|_{r'=\hat r(r),t'=\bar t'(\hat r(r))}+
\ls\dd_{t'}\ti\ff'(t',r')-\dd_{r'}\ti\ff'(t',r')\rs|_{r'=r,t'=\bar t'(r)}=0.
\nom}
Note that the relation \eqref{kr6} has a straightforward physical interpretation: $\bar t'(r)-\bar t'(\hat r(r))$ represents the time in Lorentzian coordinates taken by a spherical wave to travel from radius $\hat r(r)$ to the origin and then to radius $r$, and this time equals $\hat r(r)+r$.
\cv{As a result, it can be said that if some converging spherical wave crosses the shell at a radius \(\hat r(r)\), then after reflection from the origin, it will cross the shell again at a radius \(r\).
}

The satisfaction of condition \eqref{kr7} is necessary but not sufficient for the solution of equation \eqref{kr3} to take the form of \eqref{kr4} inside the shell. The reason is that obtaining \eqref{kr5} involves differentiating equation \eqref{kr4}. It is evident that if the form of the field \eqref{kr4} is replaced by a slightly more general form:
\disn{kr8}{
\ti\ff'(t',r')=a(t'+r')-a(t'-r')+Q,
\nom}
where $Q$ is some constant, then the condition on the shell \eqref{kr7} will still be satisfied. Furthermore, it can be proven that \eqref{kr7} becomes both a necessary and sufficient condition for the solution of equation \eqref{kr3} to take the form of \eqref{kr8} inside the shell.

Since $Q = \ti\ff'(t',0)$, a nonzero value of this quantity physically implies the existence of a delta-like source for the field $\ff'$ at the origin
(see after \eqref{kr3.1}). In this case, it turns out that we have found a solution not for the original equation \eqref{vv1}, but for the more general equation \eqref{vv2}. To find a solution specifically for the original equation \eqref{vv1}, it is necessary, in addition to satisfying the shell condition \eqref{kr7}, to impose an additional condition:
\disn{kr8.2}{
Q=\ti\ff'(t',0)=0.
\nom}
If needed, the additional condition \eqref{kr8.2} can also be formulated as conditions on the values of the field and its derivatives on the shell, but such a condition is sufficiently complex. We will limit ourselves to using the simpler condition \eqref{kr7} and separately analyze at the end whether $Q$ can be made zero.

Note that for a scalar field source (unlike, for example, a charge -- a source in electrodynamics), there is no conservation law, so, in general, $Q$ could depend on time. However, it turns out that the shell condition \eqref{kr7} guarantees that $Q = \text{const}$.

The condition on the shell \eqref{kr7}, arising from the form of the solution inside the shell, can be considered as a boundary condition when solving equations outside the shell. To do this, we first need to understand how it is formulated in terms of the values of the field and its first derivatives, interpreted as limits from outside the shell. For this, we first rewrite \eqref{kr7} in terms of $\ff'(x')$ instead of $\ti\ff'(x')$ and then transition from the coordinates $t', r'$ to $t, r$, resulting in an equation for $\ff(x)$:
\disn{kr9}{
\!\!\!\!\!\!\!\!\left.\ls r\frac{\dd t}{\dd t'}\dd_t\ff(t,r)+ r\frac{\dd r}{\dd t'}\dd_r\ff(t,r)+
r\frac{\dd t}{\dd r'}\dd_t\ff(t,r)+r\frac{\dd r}{\dd r'}\dd_r\ff(t,r)+\ff(t,r)\rs
\right|_{r=\hat r(\ti r),t=\bar t(\hat r(\ti r))}+\ns+
\left.\ls r\frac{\dd t}{\dd t'}\dd_t\ff(t,r)+ r\frac{\dd r}{\dd t'}\dd_r\ff(t,r)-
r\frac{\dd t}{\dd r'}\dd_t\ff(t,r)-r\frac{\dd r}{\dd r'}\dd_r\ff(t,r)-\ff(t,r)\rs
\right|_{r=\ti r,t=\bar t(\ti r)}=0,\!\!\!\!
\nom}
which must hold for all $\ti r > R$. Since the field $\ff$ must be continuous on the shell along with its first derivatives, the values in condition \eqref{kr9} can be understood not only as limits from within the shell (as originally considered) but also as limits from outside the shell. The partial derivatives of unprimed coordinates with respect to primed coordinates in \eqref{kr9} can be determined by inverting the matrix of partial derivatives of primed coordinates with respect to unprimed coordinates, the method for which is described after \eqref{kl7}.

As a result, solving equation \eqref{kr1} outside the shell with the boundary condition \eqref{kr9} accounts for both the equations inside the shell and the matching conditions on it. The only remaining task, if we want to solve the original equation \eqref{vv1} rather than the more general equation \eqref{vv2}, is to additionally enforce condition \eqref{kr8.2}.

\begin{figure}[htbp]\centering
\includegraphics[height=0.35\textwidth]{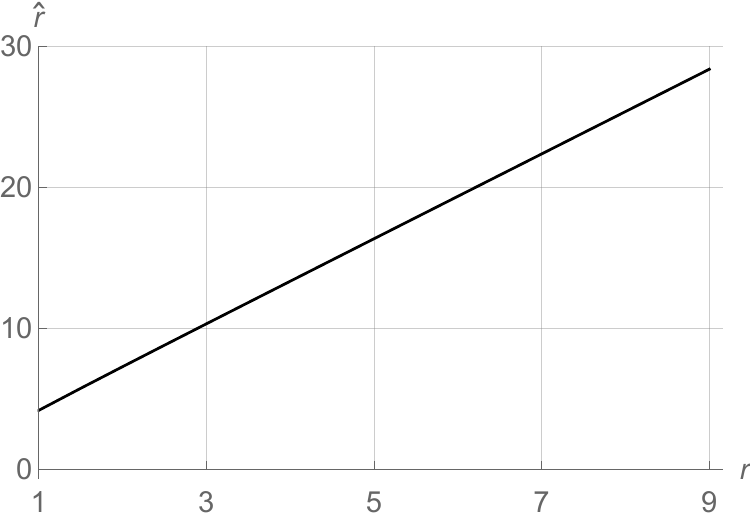}
\caption{Dependence of $\hat r$ on $r$ in units of $R$, as determined by relation \eqref{kr12}.}
\label{rkr}
\end{figure}

Before moving on to solving the equation outside the shell, let us discuss how to determine the values of $\hat r(r)$ defined by relation \eqref{kr6} for a given $r$. First, we transition in this relation from primed to unprimed coordinates (noting that on the shell $r' = r$):
\disn{kr10}{
t'(\bar t(\hat r),\hat r)+\hat r=t'(\bar t(r),r)-r.
\nom}
This equation can be rewritten as
\disn{kr11}{
-\int\limits_r^{\hat r}\!dr\frac{d}{d r}t'(\bar t(r),r)
= \hat r+r,
\nom}
which, using \eqref{kl8}, leads to the relation
\disn{kr12}{
\int\limits_r^{\hat r}\!dr
\ls\left.\ls-\frac{1}{v(r)}\frac{\dd t'}{\dd t}-
\frac{\dd t'}{d r}\rs\right|_{t=\bar t(r)}\rs
= \hat r+r.
\nom}
The integrand here can be determined explicitly for a specific form of the velocity $v(r)$ given in \eqref{kl10}. Although the integral cannot be expressed in elementary functions, the dependence $\hat r(r)$ can be computed numerically. Its graph in units of $R$ is shown in Fig.~\ref{rkr}.
It turns out that $\hat r(r) \approx 3r$ as $r \to \infty$ and $\hat r(R) = 4.14R$.

\section{Analysis of the equation outside the shell}\label{razd-sn}
We now analyze equation \eqref{kr1} outside the shell, where the metric corresponds to \eqref{kl1}. Since we are interested in possible radiation that can be detected by a distant observer located outside the horizon, it suffices to consider only the region of spacetime outside the shell and beyond the horizon, i.e., the shaded area shown in Fig.~\ref{penr}. Under the assumption of spherical symmetry of the field, the equation takes the form
\disn{sn1}{
\dd^2_{t}\ff-\frac{1}{r^2}\ls 1-\frac{R}{r}\rs\dd_{r}\ls r^2\ls 1-\frac{R}{r}\rs\dd_r\ff\rs=0.
\nom}
For further analysis, it is convenient to again introduce the field $\ti\ff \equiv r\ff$, as was done when analyzing the equation inside the shell (see before \eqref{kr2}). Using the well-known tortoise coordinate
\disn{sn1.1}{
\chi=r+R\log\frac{r-R}{R},
\nom}
and rewriting the equation in terms of the field $\ti\ff$, we obtain
\disn{sn2}{
\dd^2_{t}\ti\ff-\dd^2_{\chi}\ti\ff+\frac{R}{r^3}\ls 1-\frac{R}{r}\rs\ti\ff=0,
\nom}
where the radius \( r \) is expressed in terms of \( \chi \). Note that the generalization of equation \eqref{sn2} to the more general case of a massive field and the presence of non-zero spherical harmonics can be found in \cite{akhmhawk}. Performing a Fourier decomposition in time
\disn{sn3}{
\ti\ff(t,\chi)=\int\! d\om\, e^{i\om t}\ti\ff_\om(\chi),
\nom}
equation \eqref{sn2} can be rewritten as
\disn{sn4}{
\dd^2_{\chi}\ti\ff_\om+\om^2\ti\ff_\om-\frac{R}{r^3}\ls 1-\frac{R}{r}\rs\ti\ff_\om=0.
\nom}
The potential in this equation is similar to the well-known Regge-Wheeler potential \cite{rege-uiler}.

It can be shown that for \( \chi \to \pm\infty \) (when \( r \) approaches infinity or \( R \)), the last term in this equation can be neglected. As a result, for \( \chi \to \infty \) (noting that a similar approach applies for \( \chi \to -\infty \)), the general solution of equation \eqref{sn4} can be written as
\disn{sn5}{
\ti\ff_\om=A_\om\ti\ff^+_\om+B_\om\ti\ff^-_\om,
\nom}
where \( \ti\ff^\pm_\om \) are two solutions uniquely defined by their asymptotics
\disn{sn6}{
\ti\ff^\pm_\om-e^{\pm i\om\chi}\str{\chi\to\infty}0
\nom}
(so that \( \ti\ff^\pm_\om \) are dimensionless). It is easy to see that these two solutions are related to each other and to the results of their complex conjugates. Introducing the notation \( q_{\om} \equiv \ti\ff^-_\om \), we can write
\disn{sn6.1}{
\ti\ff^+_\om=q_{-\om},\qquad
\ti\ff^+_\om{}^*=q_{\om},\qquad
q_{-\om}=q_{\om}^*
\nom}
(where \( * \) denotes complex conjugation), from which \( \ti\ff^\pm_\om \) can be expressed in terms of \( q_{\om} \) with \( \om \geq 0 \). As a result, the general solution of equation \eqref{sn2}, i.e., the field equation outside the shell, can be written as
\disn{sn7}{
\ti\ff(t,\chi)=\int\! d\om\, \ls
A_\om e^{i\om t}q_{\om}^*(\chi)+B_\om e^{i\om t}q_{\om}(\chi)\rs.
\nom}
The first term corresponds to a solution that, at large \( r \), appears as a wave converging toward the center, while the second term corresponds to a wave diverging from the center.

Note that the real-valued nature of the field under consideration leads to the relations
$A_\om^* = A_{-\om}$ and $B_\om^* = B_{-\om}$. In quantum theory, from the perspective of an observer located at large $r$, the coefficients $A_\om$ act as creation and annihilation operators (for $\om > 0$ and $\om < 0$, respectively) for "in"-states describing the quantum system in the past, while the coefficients $B_\om$ serve the same role for "out"-states describing it in the future. Incorporating the boundary condition on the shell \eqref{kr9} obtained in the previous section introduces a relationship between the coefficients $A_\om$ and $B_\om$. For quantum theory, this relationship provides a pathway to derive an expression for the $S$-matrix; however, in this work, we limit ourselves to considering only the classical theory.

We are interested in solutions of equation \eqref{sn2} with the boundary condition \eqref{kr9}, where $A_\om = 0$, i.e., there are no converging waves, but diverging waves may be present. If such a solution exists, it would correspond to the presence of a classical \cv{version} of Hawking radiation. Rewriting the boundary condition \eqref{kr9} in terms of the field $\ti\ff$ (recall that $\ti\ff \equiv r\ff$), and then using the expression \eqref{sn7} for this field under the assumption $A_\om = 0$, we obtain the boundary condition on the shell in the form
\disn{sn8}{
\int\! d\om\, S_\om(r) B_\om = 0,
\nom}
where we introduce the notation
\disn{sn9.1}{
S_\om(r) = \frac{r-R}{r}
\ls S^+_\om(\hat r(r)) + S^-_\om(r) \rs,
\nom}
\vskip -1em
\disn{sn9}{
S^\pm_\om(r) = e^{i\om t}
\left.\ls \ls\frac{\dd t}{\dd t'} \pm \frac{\dd t}{\dd r'}\rs i\om q_\om
+\ls\frac{\dd r}{\dd t'} \pm \frac{\dd r}{\dd r'}\rs \dd_r q_\om
+\ls \pm 1 - \frac{\dd r}{\dd t'} \mp \frac{\dd r}{\dd r'}\rs \frac{1}{r} q_\om \rs
\right|_{t=\bar t(r)}.
\nom}
The condition \eqref{sn8} must be satisfied at all points along the world line of the shell outside the horizon (as stated at the beginning of this section, we are only interested in the solution outside the horizon), i.e., for $r > R$. Thus, \eqref{sn8} represents an integral equation for the function $B_\om$, which depends on the frequency $\om$. Finding this function is the main task of this work, as it allows determining the field values detected by a distant observer. Note that in the definition \eqref{sn9.1}, the factor $(r-R)/r$ is not strictly necessary since it is always nonzero in the region of interest, $r > R$. It is introduced for convenience to eliminate the pole present in $S^-_\om(r)$ as $r \to R$.

To determine the quantities \eqref{sn9} that define the kernel $S_\om(r)$ of the integral equation \eqref{sn8}, one first needs to compute the partial derivatives appearing there. This is straightforward, as the method was discussed after \eqref{kr9}. Secondly, and more challenging, one must calculate the function $q_\om$, which is a solution to equation \eqref{sn4} with the asymptotic behavior
\disn{sn10}{
q_\om - e^{-i\om\chi}\str{\chi\to\infty}0.
\nom}

To solve this problem, it is convenient to consider equation \eqref{sn1} in terms of the function $\ff$, instead of \eqref{sn4}. Performing a Fourier decomposition in time similar to \eqref{sn3} for the function $\ff$, equation \eqref{sn1} can be rewritten as
\disn{snh1}{
\frac{1}{r^2}\ls 1-\frac{R}{r}\rs\dd_{r}\ls r^2\ls 1-\frac{R}{r}\rs\dd_r\ff_\om\rs+\om^2\ff_\om=0.
\nom}
Now, substituting $\ff_\om = e^{-i\om\chi}h_\om(r)$ (where, as always, $\chi$ and $r$ are related by \eqref{sn1.1}), we obtain an equation for $h_\om$ in the form
\disn{snh2}{
r(r-R)h''_\om+(-R+2r-2i\om r^2)h'_\om-2i\om r h_\om=0.
\nom}
This equation can be reduced to the so-called confluent Heun differential equation
\cv{
\disn{snh2a}{
H''(r)+\ls \frac{\gamma}{r}+\frac{\de}{r-1}+\epsilon\rs H'(r)+\frac{\al r-\beta}{r(r-1)} H(r)=0.
\nom}
This second-order linear equation is well studied, see, for example, the monograph \cite{slav}. It has regular singularities at \(z = 0\) and \(z = 1\), and an irregular singularity of rank 1 at \(z = \infty\) \cite{slav}. The solution to this equation can be expressed in terms of the confluent Heun function \(H_c(\beta, \alpha, \gamma, \delta, \epsilon; r)\). Algorithms for the numerical computation of this function are available in various computer algebra systems, see also \cite{motygin}. Thus,} $h_\om$ can be computed numerically.

Equation \eqref{snh1} has two linearly independent solutions. Since the confluent Heun function is defined under an additional subsidiary condition, the function $\ff_\om = e^{-i\om\chi}h_\om(r)$ derived from it provides one solution to equation \eqref{snh1}. To find the second solution, note that the quantity $\om$ enters \eqref{snh1} only as $\om^2$, implying that $\ff_\om = e^{i\om\chi}h_{-\om}(r)$ is also a solution of \eqref{snh1}. By taking a linear combination of these two solutions, the desired function $q_\om$ can be sought in the form (recalling that $\ti\ff \equiv r\ff$):
\disn{snh3}{
q_\om=\xi^+_\om e^{-i\om\chi}r h_\om(r)+\xi^-_\om e^{i\om\chi}r h_{-\om}(r),
\nom}
where the coefficients $\xi^\pm_\om$, which depend only on $\om$, must be determined from the condition \eqref{sn10} for each value of $\om$. Since the values of the function $h_\om$ can only be computed numerically, the coefficients $\xi^\pm_\om$ must also be determined numerically. The most convenient method for their numerical determination is described in Appendix~2. This method uses a more precise asymptotic behavior of $q_\om$ at large $r$ compared to \eqref{sn10}, which will be derived in the next section.

Thus, we have the capability to numerically compute the kernel $S_\om(r)$ of the integral equation \eqref{sn8} for various values of $\om$ and $r$. Consequently, solving this equation can also only be done numerically, which requires discretizing the parameters $\om$ and $r$ and imposing some bounds on them. As a result, the problem reduces to finding the eigenvector of a matrix corresponding to a zero eigenvalue. Introducing truncations for $\om$ and $r$ can significantly distort the problem, causing computations performed with finite truncations to deviate from the exact solution. It is necessary to find a method of introducing truncations that makes the matrix finite without introducing such distortions. This is the focus of the next section.

\section{Asymptotics of the quantities under study}\label{razd-as}
We will demonstrate how various asymptotics for the solutions of equation \eqref{sn4} can be determined with sufficiently high accuracy. This equation can be rewritten as
\disn{as1}{
\dd^2_{\chi}\ti\ff_\om+\om^2\ti\ff_\om-V(r)\ti\ff_\om=0,\qquad
V(r)\equiv\frac{R}{r^3}\ls 1-\frac{R}{r}\rs.
\nom}
The introduced function \( V(r) \) is positive (we consider only \( r > R \)), approaches zero as \( r \to \infty \) or \( r \to R \), and is bounded above by \( V_{\text{max}} = 27/(256R^2) \). This implies that for \( \om^2 \gg V_{\text{max}} \), the last term in equation \eqref{as1} can also be neglected (as is the case when \( r \to \infty \) or \( r \to R \)), in comparison to the contribution of \( \om^2\ti\ff_\om \). As a result, we can seek solutions to this equation in the specified asymptotic regimes using perturbation theory. We take \( e^{\pm i\om\chi} \) as the zeroth-order solution and introduce a correction to it by writing the solution as
\disn{as2}{
e^{\pm i\om\chi}\ls 1+\si_\om\rs,
\nom}
where the function \( \si_\om(\chi) \) is small. Substituting this into equation \eqref{as1}, we obtain
\disn{as3}{
\dd^2_{\chi}\si_\om \pm 2i\om\dd_{\chi}\si_\om-V(r)=0,
\nom}
where the term \( -V(r)\si_\om \) is neglected as it is small compared to \( -V(r) \).

First, consider the case where the last term in \eqref{as1} can be neglected because either \( \om^2 \gg V_{\text{max}} \) or \( r \gg R \). In this situation, the first term in \eqref{as3} can also be neglected compared to the second term, as the latter contains the factor \( \om \), while the former involves an additional derivative with respect to \( \chi \), which leads to faster decay for the power-law behavior that \( \si_\om \) exhibits at large \( r \). Using \( \dd_\chi = (1-R/r)\dd_r \) and the expression for \( V(r) \), we find \( \si_\om \), and the solution \eqref{as2} in the first approximation in the specified asymptotics becomes
\disn{as4}{
e^{\pm i\om\chi}\ls 1\mp \frac{R}{4i\om r^2}\rs.
\nom}
For the function \( q_\om \), which is part of the kernel \eqref{sn9.1} of the integral equation \eqref{sn8} and satisfies the main asymptotics \eqref{sn10}, this yields the asymptotic expression
\disn{as5}{
q_\om\approx e^{-i\om\chi}\ls 1+\frac{R}{4i\om r^2}\rs
\nom}
for \( |\om| \to \infty \) or \( r \to \infty \).

Next, let us consider the case where the last term in \eqref{as1} can be neglected because \( r \to R \) (recall that in this case, \( \chi \to -\infty \)). The potential \( V(r) \) can then be approximated using \eqref{sn1.1} as \( e^{-1+\chi/R}/R^2 \). Substituting this into \eqref{as3}, we find a solution for \( \si_\om \) of the form
\disn{as6}{
\si_\om=\frac{e^{\frac{\chi}{R}-1}}{1\pm 2i\om R}.
\nom}
As a result, the solution \eqref{as2} in the first approximation in the considered asymptotic regime becomes
\disn{as7}{
e^{\pm i\om\chi}\ls 1+\frac{e^{\frac{\chi}{R}-1}}{1\pm 2i\om R}\rs.
\nom}
The function \( q_\om \), whose asymptotics at \( \chi \to \infty \) are fixed by condition \eqref{sn10}, will at \( \chi \to -\infty \) be approximately described by a linear combination of the expressions in \eqref{as7} with coefficients depending on \( \om \):
\disn{as8}{
q_\om\approx
K^+_\om
e^{i\om\chi}\ls 1+\frac{e^{\frac{\chi}{R}-1}}{1+ 2i\om R}\rs+
K^-_\om
e^{-i\om\chi}\ls 1+\frac{e^{\frac{\chi}{R}-1}}{1- 2i\om R}\rs.
\nom}
The determination of the coefficients \( K^\pm_\om \) is possible only numerically, and this task shares similarities with finding \( \xi^\pm_\om \) in the decomposition \eqref{snh3} for \( q_\om \). The numerical method for determining \( K^\pm_\om \) is described in Appendix~3, where graphs of the computed results are also provided.

It is worth noting that, given the property \( q_\om^* = q_{-\om} \) (see \eqref{sn6.1}), a similar property holds for the coefficients introduced here: \( K^\pm_\om{}^* = K^\pm_{-\om} \). Additionally, if \( \om \to \infty \) and \( \chi \to -\infty \), both asymptotics \eqref{as5} and \eqref{as8} must simultaneously hold. This is possible if \( K^+_\om \to 0 \) and
\disn{as8.1}{
K^-_\om\approx 1+\frac{1}{4i\om R}
\nom}
as \( \om \to \infty \). This behavior of the coefficients \( K^\pm_\om \) is confirmed by numerical computations (see Fig.~\ref{grafK} in Appendix~3).

Using the obtained asymptotics for \( q_\om \), one can derive the corresponding asymptotics for the kernel \( S_\om(r) \) in \eqref{sn9.1}. To simplify numerical analysis, we use a parameter \( \al \) that is monotonically related to \( r \) or the tortoise coordinate \( \chi \) (see \eqref{sn1.1}):
\disn{as9}{
\al=3r+R\log\frac{2(r-R)^2}{R(r+R)}-R.
\nom}
Note that as \( r \to R \), \( \al \), like \( \chi \), tends to \( -\infty \), with \( \al \sim 2\chi \).

For \( |\om| \to \infty \) or \( r \to \infty \), substituting \eqref{as5} into \eqref{sn9.1} shows that in these limiting cases, \( S_\om \approx \bar S_\om \), where
\disn{as10}{
\bar S_\om(\al)=e^{-i\om\al}\Bigl( f_1(r(\al))+i\om f_2(r(\al))\Bigr)+e^{-i\om\hat\al(\al)}\frac{r(\al)-R}{r(\al)}f_3(\hat r(r(\al))).
\nom}
Here, \( \hat\al(\al) \equiv \al(\hat r(r(\al))) \), where \( \al(r) \) is the relationship given in \eqref{as9}, \( r(\al) \) is its inverse, and \( \hat r(r) \) is the dependence defined by \eqref{kr12}. The functions \( f_{1,2,3}(r) \) are explicitly determined and listed in Appendix~4. The agreement between \( S_\om \) and \( \bar S_\om \) holds with a relative accuracy of 0.1\% for all \( \al \), provided \( |\om| > 6/R \), as well as for \( \al > 30 \) when \( |\om| > 0.1/R \) (for very small \( |\om| \), the accuracy decreases).

For \( r \to R \) (and thus \( \al \to -\infty \)) at arbitrary \( \om \), substituting \eqref{as8} into \eqref{sn9.1} shows that in this limit
\disn{as11}{
S_\om\approx
4\ls\sqrt{2}-1\rs i\om K^-_\om e^{-i\om\al}.
\nom}
This asymptotics is accurate to within 0.1\% for any \( \om \) if \( \al < -17 \).

\section{Reduction of the integral equation to a matrix form}\label{razd-ob}
We aim to develop an accurate and structured approach to reduce the integral equation \eqref{sn8} to a matrix equation that can be solved numerically. To achieve this, the quantity \(B_{\om}\) is expressed in terms of its Fourier transform
\disn{ob1}{
B_\om=\int\! d\ti\al\, e^{i\om\ti\al}B(\ti\al).
\nom}
Substituting this into \eqref{sn8} transforms the integral equation into
\disn{ob2}{
\int\! d\ti\al\, S(\al,\ti\al)B(\ti\al)=0,
\nom}
where the kernel \( S(\al, \ti\al) \) is defined as
\disn{ob3}{
S(\al,\ti\al)=\int\! d\om\, S_\om(\al)e^{i\om\ti\al}.
\nom}
This form of the integral equation \eqref{ob2} is more convenient because both arguments of the kernel now have the same nature, unlike in the original equation \eqref{sn8}.

We next compute a similar expression for the asymptotic kernel \( \bar S_\om(\al) \) given in \eqref{as10}:
\disn{ob4}{
\bar S(\al,\ti\al)=\int\! d\om\, \bar S_\om(\al)e^{i\om\ti\al}=\ns=
2\pi\ls
f_1(r(\al))\de(\ti\al-\al)+f_2(r(\al))\de'(\ti\al-\al)+
\frac{r(\al)-R}{r(\al)}f_3(\hat r(r(\al)))\de(\ti\al-\hat\al(\al))\rs.
\nom}
The integral equation \eqref{ob2} can now be rewritten as
\disn{ob5}{
\int\! d\ti\al\, \ls \bar S(\al,\ti\al)+\ti S(\al,\ti\al)\rs B(\ti\al)=0,
\nom}
where the deviation \( \ti S(\al,\ti\al) \) is defined as
\disn{ob6}{
\ti S(\al,\ti\al)=S(\al,\ti\al)-\bar S(\al,\ti\al)=
\int\! d\om\, \ls S_\om(\al)-\bar S_\om(\al)\rs e^{i\om\ti\al}.
\nom}
From the previous section (see before \eqref{as10}), it is evident that \( S_\om \approx \bar S_\om \) for large \( |\om| \). Therefore, the integration in \eqref{ob6} can be truncated with \( |\om| < \om_{\text{max}} \), introducing only a small error if \( \om_{\text{max}} \) is sufficiently large. Additionally, using \( S^*_\om = S_{-\om} \) and \( \bar S^*_\om = \bar S_{-\om} \), we find the approximate relation:
\disn{ob7}{
\ti S(\al,\ti\al)\approx
2\int\limits_0^{\om_{\text{max}}}\! d\om\, \text{Re}\Bigl(\ls S_\om(\al)-\bar S_\om(\al)\rs e^{i\om\ti\al}\Bigr),
\nom}
which becomes increasingly accurate as \( \om_{\text{max}} \to \infty \).

\begin{figure}[htbp]\centering
\includegraphics[height=0.35\textwidth]{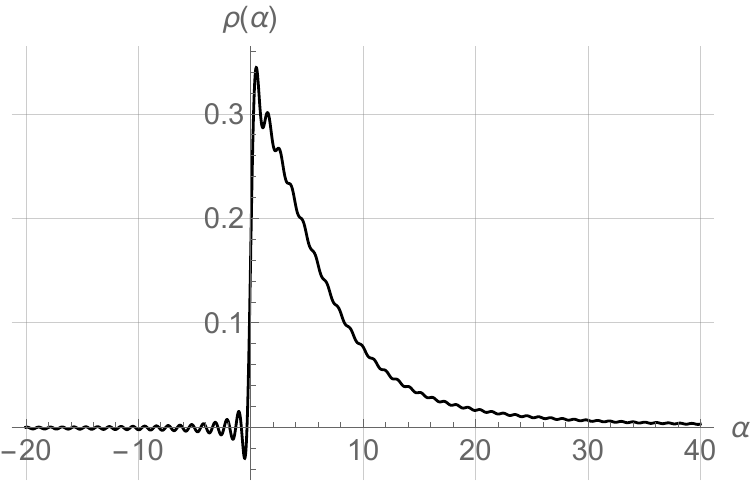}
\caption{The graph of the dimensionless function $\rho(\al)$, defined in \eqref{ob9}.
The horizontal axis represents the values of $\al$ in units of $R$.}
\label{funrho}
\end{figure}

Let us analyze the behavior of contributions to the integral equation \eqref{ob5} when \( \al \to -\infty \). Using the asymptotic limits provided in Appendix~4 (see \eqref{prf4}), we deduce that as \( \al \to -\infty \) (i.e., \( r \to R \)):
\disn{ob8}{
\bar S(\al,\ti\al)\approx
2\pi\ls\sqrt{2}-1\rs \ls\frac{1}{R}\de(\al-\ti\al)-4\de'(\al-\ti\al)\rs.
\nom}
For \( \ti S(\al,\ti\al) \), substituting the asymptotics \eqref{as11} for \( S_\om \) into \eqref{ob6} and using the limits from \eqref{prf4} in \eqref{as10} for \( \bar S_\om(\al) \), we obtain
\disn{ob9}{
\ti S(\al,\ti\al)\approx \frac{1}{R^2}\rho(\ti\al-\al),\qquad
\rho(\al)=
4\ls\sqrt{2}-1\rs R^2\int\! d\om\, e^{i\om\al} i\om \ls  K^-_\om -1-\frac{1}{4i\om R}\rs.
\nom}
Taking into account the asymptotics \eqref{as8.1}, it can be noted that the function $\rho(\al)$ does not contain $\de$-functional contributions. The graph of this function is shown in Fig.~\ref{funrho}.

Now, let us examine how the contributions to \eqref{ob5} behave as $\al\to\infty$. Using the limits \eqref{prf5} presented in Appendix~4 in equation \eqref{ob4}, it can be concluded that for $\al\to\infty$ (i.e., as $r\to\infty$), we have
\disn{ob10}{
\bar S(\al,\ti\al)\approx
4\pi\de'(\ti\al-\al).
\nom}
For the quantity $\ti S(\al,\ti\al)$ in this limit, we find
\disn{ob10.1}{
\ti S(\al,\ti\al)\str{\al\to\infty}0,
\nom}
as discussed above and below equation \eqref{as10}.

The asymptotics obtained also allow us to present some considerations about the behavior of the solution $B(\al)$ of the integral equation \eqref{ob5} in the limits $\al\to\pm\infty$ even before performing calculations. For $\al\to-\infty$, equation \eqref{ob5}, taking into account the asymptotics \eqref{ob8} and \eqref{ob9}, takes the form
\disn{obn11}{
4R B'(\al)-B(\al)-\frac{1}{2\pi\ls\sqrt{2}-1\rs R}
\int\! d\ti\al\,\rho(\ti\al-\al)B(\ti\al)=0.
\nom}
The function $\rho(\al)$, in some approximation, has a bounded support and is positive within it, see Fig.~\ref{funrho}. Therefore, if the solution $B(\al)$ does not vary too rapidly at large negative $\al$, the last term in \eqref{obn11} can roughly be replaced by $C B(\al)$, where $C$ is a certain positive dimensionless constant (calculations yield a value close to unity). As a result, the equation simplifies to
\disn{obn12}{
4R B'(\al)-(1+C) B(\al)=0.
\nom}
Its solution tends to zero as $\al\to-\infty$. As will be shown below, numerical calculations confirm this asymptotic behavior of the solution $B(\al)$.

Conversely, if we consider equation \eqref{ob5} for $\al\to\infty$ and take into account \eqref{ob10} and \eqref{ob10.1}, the equation reduces, to leading order, to the condition $B'(\al)=0$. This condition is consistent with the asymptotic behavior $B(\al)\str{\al\to\infty}const$, although the possibility of corrections to this behavior, such as logarithmic ones, cannot be ruled out. Below, we will provide additional arguments in favor of the constant asymptotic behavior. Numerical results do not contradict the constant asymptotic behavior of the solution $B(\al)$ for large positive $\al$.

To numerically solve the integral equation \eqref{ob5} by reducing it to a matrix equation with a finite matrix size, it is necessary to introduce truncation on both sides for the independent variable $\al$ and the integration variable $\ti\al$. As a result, we transition to the equation
\disn{ob13}{
\int\limits_{\ti\al_1}^{\ti\al_2}\! d\ti\al\, \ls \bar S(\al,\ti\al)+\ti S(\al,\ti\al)\rs B(\ti\al)=0,
\nom}
which is assumed to be valid only within $\al_1<\al<\al_2$.

If $\al_2>\ti\al_2$ is chosen, then considering \eqref{ob10}, \eqref{ob10.1}, the equation for $\ti\al_2<\al<\al_2$ will lose a significant (delta-functional) contribution, leading to significant distortion of the result. A similar issue arises if $\al_1<\ti\al_1$, as delta-functional contributions corresponding to \eqref{ob8} will be lost.

On the other hand, if $\al_2<\ti\al_2$, various solutions $B(\ti\al)$ with support within $\al_2<\ti\al<\ti\al_2$ emerge. These are clearly unphysical, as they would be prohibited when considering equations with $\al>\al_2$. A similar issue arises if $\al_1>\ti\al_1$, though it is slightly less obvious due to the presence of not only delta-functional contributions: for $\al\to-\infty$, the contribution of $\ti S(\al,\ti\al)$ \eqref{ob9} is non-zero (unlike the contribution \eqref{ob10.1} for $\al\to\infty$). However, as seen from Fig.~\ref{funrho}, the function $\rho(\al)$ decreases sharply for $\al<0$. Numerical calculations confirm the existence of this problem  --  unphysical solutions in this sense arise.

As a result, we conclude that to obtain correct results, the truncations for $\al$ and $\ti\al$ must be introduced consistently. We set
\disn{ob14}{
\ti\al_2=\al_2, \qquad \ti\al_1=\al_1.
\nom}
It is worth noting that the introduced truncation is justified since, for $\al\to-\infty$, considering \eqref{ob8}, \eqref{ob9}, and the form of the function $\rho(\al)$, the matrix is nearly upper triangular. For $\al\to\infty$, considering \eqref{ob10}, \eqref{ob10.1}, the matrix is nearly diagonal. Moreover, for fixed $\al$, the value of $S(\al,\ti\al)$ \eqref{ob3} decreases as $\ti\al\to\pm\infty$ due to the continuity of $q_\om$ in $\om$, and hence of $S_\om(r)$ \eqref{sn9.1}.

To reduce the integral equation \eqref{ob13} to a matrix form, the variables $\al$ and $\ti\al$ must be discretized. We assume they take discrete values with a step size $\De\al$. The delta function and its derivative, appearing in the definition \eqref{ob8} for $\bar S(\al,\ti\al)$, will also be discretized in a standard manner. Additionally, we introduce the discretization of the parameter $\om$ in the formula defining $\ti S(\al,\ti\al)$ \eqref{ob7}, which converts the integral into a sum. The step size in $\om$ is denoted as $\De\om$. As a result, equation \eqref{ob13} is reduced to the matrix equation
\disn{ob15}{
\ls \bar S+\ti S\rs B=0,
\nom}
where $\bar S$ and $\ti S$ are matrices obtained from the discretization of $\bar S(\al,\ti\al)$ and $\ti S(\al,\ti\al)$, respectively, and $B$ is the vector corresponding to the desired function $B(\ti\al)$. Considering \eqref{ob14} and the need to discretize the derivative of the delta function in $\bar S(\al,\ti\al)$, either non-square matrices with one more column than rows must be used, or the last row of the matrix $\bar S$ must be filled with zeros  --  in the case of square matrices. The results are identical, but the latter method is technically more convenient.

Through the introduction of these truncations and discretizations, we obtain the corresponding parameters: $\De\om$, $\om_{\text{max}}$, $\De\al$, $\al_1$, $\al_2$. To achieve larger values of $\al_2$ without significantly increasing computations, we can use \eqref{ob10.1} and introduce an additional truncation by setting $\ti S(\al,\ti\al)=0$ for $\al_3<\al<\al_2$. Then, it suffices to calculate $\ti S(\al,\ti\al)$, which depends on Heun functions, only for $\al<\al_3$, allowing $\al_2$ to be chosen much larger, since $\bar S(\al,\ti\al)$ is defined by the relatively simple expression \eqref{ob4}.

\section{Results}\label{razd-rz}
Ultimately, the numerical solution of equation \eqref{ob15} reduces to finding the eigenvector of the matrix $\bar S+\ti S$ corresponding to the zero eigenvalue. The calculations show that this eigenvector is unique, as all other eigenvalues of the matrix $\bar S+\ti S$ are significantly separated from zero, and this fact remains unchanged when the truncation and discretization parameters are varied.

\begin{figure}[htbp]\centering
\includegraphics[height=0.4\textwidth]{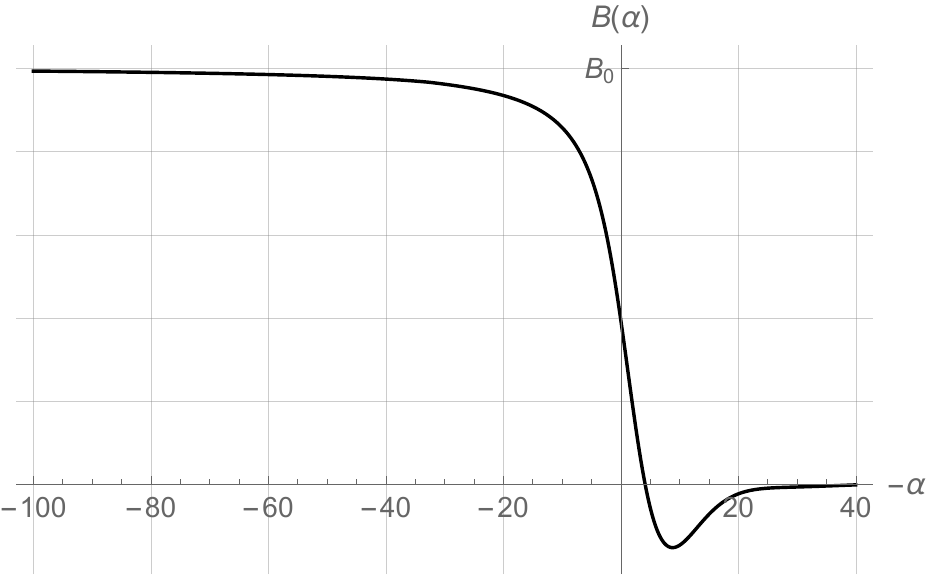}
\caption{The graph of the solution $B(\al)$ as a function of $-\al$, plotted in units of $R$.
According to \eqref{rz3}, the same graph represents the variation of the field $\ti\ff(t,\chi)$,
as observed by a distant observer, with time $t$, assuming that the horizontal axis corresponds to $t-\chi$ in units of $R$.}
\label{grafB}
\end{figure}

For the calculations, we used the following
\cv{maximum (minimum) values of the truncation (discretization)}
parameters described in the previous section:
\disn{rz1}{
\De\om=\frac{1}{512R},\qquad\!
\om_{\text{max}}=\frac{6}{R},\qquad\!
\De\al=\frac{R}{50},\qquad\!
\al_1=-40R,\qquad\!
\al_2=100R,\qquad\!
\al_3=40R.
\nom}
The resulting matrix has dimensions of 7001 by 7001. Performing calculations with significantly
\cv{smaller for truncations or larger for discretizations}
parameter values shows that the results remain practically unchanged,
\cv{therefore, the values in \eqref{rz1} can be considered optimal.}
Thus, it can be concluded that within the achieved parameter range, there is no significant dependence on the parameters, and further refinement of their values is unnecessary. The graph of the solution $B(\al)$ obtained from the calculations is shown in Fig.~\ref{grafB}.
For convenience in further discussion, the horizontal axis represents $-\al$ instead of $\al$
\cv{(in this case, the graph can also be understood as the dependence of the field \(\tilde{\varphi}(t,\chi)\) on \(t-\chi\), see below).}
It should be noted that the solution is defined up to a multiplicative constant $B_0$, which could be zero -- resulting in a trivial solution. Since, as evident from the graph, $B(\al)\ne0$ as $\al\to\infty$, it is convenient to define $B_0$ as the asymptotic value of $B(\al)$ in this limit.

Given $B(\al)$, we can determine the field $\ti\ff(t,\chi)$ at any point outside the shell. Substituting \eqref{ob1} into \eqref{sn7} and assuming the absence of incoming waves ($A_{\om}=0$), we obtain:
\disn{rz2}{
\ti\ff(t,\chi)=\int\! d\al\int\! d\om\, B(\al) e^{i\om (t+\al)}q_{\om}(\chi).
\nom}
Now we can determine the values of the field observed by a distant observer at a large fixed $\chi$ (and therefore $r$). Using the asymptotics \eqref{sn6} (recall that $q_{\om}\equiv\ti\ff^-_\om$) in \eqref{rz2}, we find that for large $\chi$:
\disn{rz3}{
\ti\ff(t,\chi)=2\pi\int\! d\al\, B(\al) \de(t+\al-\chi)=2\pi B(\chi-t).
\nom}
Thus, the graph shown in Fig.~\ref{grafB} can also be interpreted as the time dependence of the field $\ti\ff(t,\chi)$ observed by a distant observer, assuming that the horizontal axis corresponds to $t-\chi$. A crucial property of the observed dependence of $\ti\ff$ on $t$ at large $\chi$ is that $\ti\ff\ne0$ as $t\to-\infty$.

As noted in Section~\ref{razd-kr} (see \eqref{kr7} and the surrounding text), the solution we found corresponds to the potential presence of a constant field source at the symmetry center. To verify whether such a source exists, we need to check the fulfillment of \eqref{kr7}, which requires calculating the constant $Q$. To do so, one must use formula \eqref{rz2} to find the values of the field $\ff(t,r)$ and its first derivatives on the shell, use the continuity of these quantities across the shell (see after \eqref{kr9}), switch to coordinates $t',r'$, and compare the result with \eqref{kr8}. This procedure ultimately yields the field inside the shell and the value of $Q$. We will not present this rather cumbersome calculation, as the same result for $Q$ can be obtained more straightforwardly.

According to \eqref{kl11}, as $t\to-\infty$, the radius of the shell grows indefinitely. In this limit, the Schwarzschild metric outside the shell \eqref{kl1} becomes indistinguishable from the Minkowski metric inside the shell \eqref{kl2}. As a result, the matching conditions on the shell become trivial, and the entire spacetime in the region $t\to-\infty$ is nearly Minkowski. In this region, assuming the absence of incoming waves, there are also no outgoing waves (note that this conclusion cannot be made a priori for later times, as outgoing waves may appear due to changes in the metric). As a result, the solution to \eqref{vv2} in the region $t\to-\infty$ reduces to a Coulomb field $\ff=Q/r$ both inside and outside the shell. Thus, in particular, for $t\to-\infty$ outside the shell (and hence as $r\to\infty$), we have $\ti\ff=r\ff=Q$. This reasoning also provides an additional argument, mentioned after \eqref{obn12}, in favor of the constant behavior of the graph in Fig.~\ref{grafB} for $\al\to\infty$. Furthermore, from \eqref{rz3}, we find that
\disn{rz4}{
Q=2\pi\lim_{\al\to\infty}B(\al)=2\pi B_0.
\nom}
The same result can be obtained through the more straightforward calculation described earlier.

It is straightforward to understand what happens if we impose condition \eqref{kr8.2}, requiring the absence of a source for the field $\ff$ and transitioning to the solution of the original field equation \eqref{vv1}. Since $B_0=0$, the only solution in this case is $B(\al)=0$, corresponding to a vanishing field according to \eqref{rz2}. Thus, within the physical scenario considered, a distant observer will not detect any radiation emitted by the black hole if the studied field has no sources.

\begin{figure}[htbp]\centering
\includegraphics[height=0.35\textwidth]{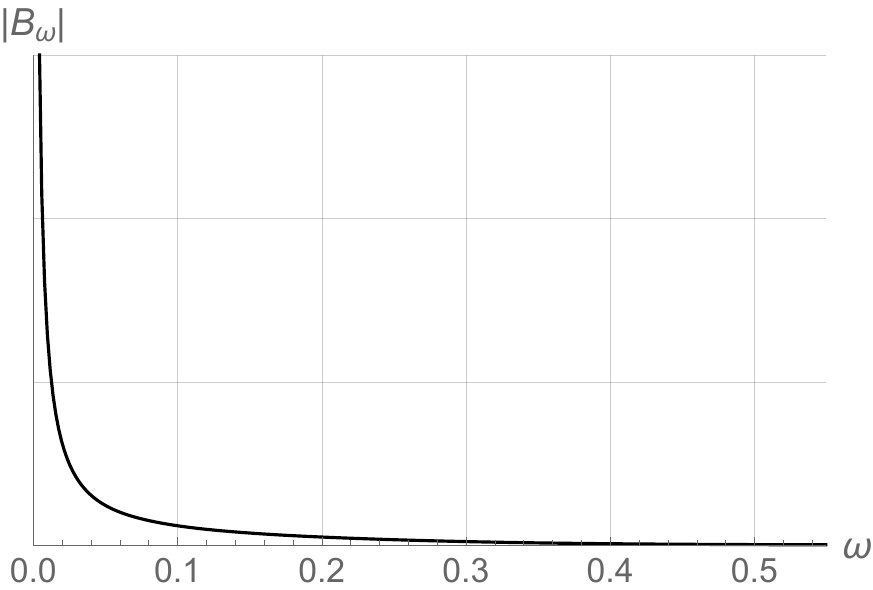}
\caption{Dependence of $|B_\om|$ on the frequency $\om$, plotted in units of $1/R$.}
\label{grafBf}
\end{figure}

In the case of a constant point source at the center of symmetry, however, the distant observer will detect a time-dependent non-zero field \eqref{rz3}. This can be interpreted as radiation emitted by the emerging black hole. The spectrum of this radiation can be found by performing a Fourier transform at $r\to\infty$:
\disn{rz5}{
\frac{1}{2\pi}\int\! dt\, e^{-i\om t}\ti\ff(t,\chi)=
\int\! dt\, e^{-i\om t} B(\chi-t)=e^{-i\om \chi}B_\om,
\nom}
where \eqref{rz3} and \eqref{ob1} are used. The graph of the modulus of this function is shown in Fig.~\ref{grafBf}.
Its asymptotics for small $\om$ is given by $|B_\om|\approx B_0/\om$.

The observed time-dependent field is shown in Fig.~\ref{grafB}. The constant behavior of $\ti\ff$ in the past, when the radius $r$ of the collapsing shell is much larger than its Schwarzschild radius $R$, corresponds to the Coulomb field of the central scalar field source. As $r$ approaches $R$ (with $t$ approaching zero for the shell, as per \eqref{kl11}, but for the distant observer, due to signal delay, this occurs when $t\approx\chi\approx r$), the value of $\ti\ff$ decreases and, after a single sign change, approaches zero. This can be interpreted as the screening of the Coulomb field by the forming black hole horizon (note that beneath the shell, where the metric is flat, the Coulomb component of the field is always observed), initially partially and then completely. It should be noted that the possibility of such screening for scalar field theory does not contradict conventional notions, unlike, for instance, in electrodynamics, where the total charge cannot be screened.

\section{Conclusion}\label{razd-za}
We attempted to construct a classical \cv{version} of Hawking radiation -- classical radiation detectable by a distant observer emitted by a black hole formed through collapse. The potential existence of such radiation arises from the non-conservation of matter energy in a non-static metric.

A specific physical scenario was considered: the collapse of a thin spherical shell contracting according to the prescribed law \eqref{kl11} (the shell's matter is not dust-like and possesses some transverse pressure). In the corresponding spacetime, spherically symmetric solutions of a real, free, massless scalar field were studied, assuming the presence of only outgoing waves, with incoming waves absent. Both solutions to the homogeneous equation \eqref{vv1}, corresponding to the absence of sources -- this formulation explores the classical \cv{version} of Hawking radiation -- and solutions to the inhomogeneous equation \eqref{vv2} with a constant point source at the center of symmetry were examined.
\cv{Such a source plays for the scalar field under consideration a similar role to that of a point charge in electrodynamics.
We introduced a point source to increase the generality of the problem and unexpectedly discovered the effect
}
of screening the source during horizon formation.

For the case without sources, it was found that no non-trivial solutions exist within the considered problem formulation. Technically, this result is tied to the fact that the limit of the function $B(\al)$ as $\al\to\infty$ is non-zero, see \eqref{rz4}. This, in turn, is due to the asymptotics \eqref{ob10} containing only a term with the derivative of the delta function but not a term with the delta function itself. If such a term were present, $B(\al)$ would tend to zero as $\al\to\infty$, and the condition $Q=0$ in \eqref{kr8.2} would no longer imply $B(\al)=0$.

It is possible that this situation may change if the problem formulation is expanded. For instance, one could consider not only spherically symmetric modes of the field but also the behavior of higher spherical harmonics. Subsequently, the study could shift from scalar fields to the more physically relevant case of electrodynamics (where, due to the transversality of the electromagnetic field, it is necessary to go beyond spherically symmetric modes). Additionally, one could investigate whether the results qualitatively change with variations in the shell's contraction law, for example, if it has a finite radius in the distant past. These extensions require further research and calculations.

For the case of a constant point source at the center of symmetry, a solution corresponding to the presence of only outgoing waves was found. A distant observer will detect this radiation in significant amounts over a finite period of time. The radiation during this period corresponds to a transition from observing the Coulomb field of the central source in the past to observing a zero field in the future.
\cv{It can be thought that the energy for the emerging radiation is provided by the initial state -- the Coulomb field of the point source, which then disappears. However, without considering the backreaction, when the energy of the scalar field alone is not conserved, this cannot be discussed strictly.
}
Since the Coulomb field is always present beneath the horizon, the latter can be considered the result of the central source being screened by the emerging horizon. Such screening is possible for scalar fields but not for electromagnetic fields. This raises the question once more of how the results obtained in this study would change if electrodynamics were considered instead of scalar field theory.

{\bf Acknowledgements.}
The authors are grateful to A.~Starodubtsev and M.~Smolyakov for useful discussions. The work of D.S.~Shatkov was supported by the Ministry of Science and Higher Education of the Russian Federation, agreement no.~075-15-2022-287.

\section*{Appendix~1}
Here, we present the explicit expressions for the partial derivatives of the Lorentzian coordinates $t',r'$ inside the shell with respect to the coordinates $t,r$, in which the metric is continuous on the shell. These values are obtained by solving the four equations \eqref{kl5}, \eqref{kl7} and are given by:
\disn{prn1}{
\frac{\dd t'}{\dd t}=\left(s\sqrt{(s-1) \left(s^2 (v(r)^2-1)+2 s-1\right)}\left( s^2(v(r)^2-1)+2 s-1\right)^{1/2}\right)^{-1}(s-1)\times \ns \times
 \left(2 s^3-s^2 \left(v(r)^2+5\right)-2 s \left(\sqrt{(s-1) s \left(s^2-s \left(v(r)^2+2\right)+1\right)}-2\right)+\right.\ns \left.+2 \sqrt{(s-1) r \left(s^2-s \left(v(r)^2+2\right)+1\right)}-1\right)^{1/2}\times \ns \times
\left(\sqrt{(s-1) r \left(s^2-s \left(v(r)^2+2\right)+1\right)}+s^2-s\right),
\nom}
\disn{prn2}{
\frac{\dd r'}{\dd t}=-sv(r)\left(\left(-1 + 2 s + s^2 (-1 + v(r)^2)\right)\left((-1 + s) (-1 + 2 s + s^2 (-1 + v(r)^2)\right)\right)^{-1/2}\times\ns \times \left(2 s^3-s^2 \left(v(r)^2+5\right)-2 s \left(\sqrt{(s-1) s \left(s^2-s \left(v(r)^2+2\right)+1\right)}-2\right)+\right. \ns \left.+2 \sqrt{(s-1) s \left(s^2-s \left(v(r)^2+2\right)+1\right)}-1\right)^{1/2},
\nom}
\disn{prn3}{
\frac{\dd t'}{\dd r}=\frac{v(r) \left(\sqrt{(s-1) s \left(s^2-s \left(v(r)^2+2\right)+1\right)}-s^2+2 s-1\right)}{s^2 \left(v(r)^2-1\right)+2 s-1},
\nom}
\disn{prn4}{
\frac{\dd r'}{\dd r}=\frac{s^2 v(r)^2-\sqrt{(s-1) s \left(s^2-r \left(v(r)^2+2\right)+1\right)}}{s^2 \left(v(r)^2-1\right)+2 s-1},
\nom}
where $s\equiv r/R$ and $v(r)$ is an arbitrary function describing the velocity of the shell in Schwarzschild coordinates as a function of its radius. For the purposes of our calculations, $v(r)$ was specified using formula \eqref{kl10}.

\section*{Appendix~2}
We describe a method for numerically determining the coefficients $\xi^\pm_\om$ in formula \eqref{snh3}, which utilizes the more accurate asymptotics \eqref{as5} for $q_\om$ as $r\to\infty$, compared to condition \eqref{sn10}. As will be shown in Appendix 3, a similar approach allows us to compute the numerical values of the coefficients $K^\pm_\om$ in formula \eqref{as8}.

Each of the functions multiplied by the coefficients $\xi^\pm_\om$ in formula \eqref{snh3} is a solution to equation \eqref{as1}. Thus, for large $r$, we can write the asymptotics for these functions, for each value of $\om$, as a linear combination of the expressions \eqref{as4} with certain coefficients, for example:
\disn{pra1}{
e^{-i\om\chi}r h_\om(r)\approx\dz^+_\om e^{i\om\chi}\ls 1-\frac{R}{4i\om r^2}\rs+
\dz^-_\om e^{-i\om\chi}\ls 1+\frac{R}{4i\om r^2}\rs.
\nom}
This can be written compactly as
\disn{pra2}{
u_\om(r)\approx\dz^+_\om v^+_\om(r)+\dz^-_\om v^-_\om(r),
\nom}
where the notation is self-explanatory. Dividing this approximate equality by $v^+_\om(r)$ or $v^-_\om(r)$ and then differentiating with respect to $r$, we find:
\disn{pra3}{
\dz^\pm_\om=\lim_{r\to\infty}\ls\frac{\dd}{\dd r}\frac{u_\om(r)}{v^\mp_\om(r)}\rs
\ls\frac{\dd}{\dd r}\frac{v^\pm_\om(r)}{v^\mp_\om(r)}\rs^{-1}.
\nom}
Since the asymptotics \eqref{pra1} include not only the leading order term but also the next order \eqref{as4}, the term under the limit in \eqref{pra3} converges sufficiently quickly, as confirmed by numerical calculations. Consequently, to obtain $\dz^\pm_\om$ with sufficient accuracy, it is enough to evaluate the term under the limit for moderately large $r$. In our calculations, we chose this value as $300/\om$. This choice is explained by the observation that the algorithms for computing the Heun function, which defines $h_\om$, work well in the region $r|\om|<300$, but may fail for larger values.

Substituting the asymptotics \eqref{pra1} into formula \eqref{snh3} (noting that to compute the contribution of the second term, it suffices to take the opposite sign of $\om$), we obtain the behavior of $q_\om$ for large $r$:
\disn{pra4}{
q_\om\approx\xi^+_\om
\ls\dz^+_\om v^+_\om(r)+\dz^-_\om v^-_\om(r)\rs
+\xi^-_\om
\ls\dz^+_{-\om} v^+_{-\om}(r)+\dz^-_{-\om} v^-_{-\om}(r)\rs.
\nom}
Comparing this asymptotics with condition \eqref{sn10}, and taking into account the definitions of $v^\pm_{-\om}(r)$, it is straightforward to derive the following relations:
\disn{pra5}{
\xi^+_\om\dz^+_\om+\xi^-_\om\dz^-_{-\om}=0,\qquad
\xi^+_\om\dz^-_\om+\xi^-_\om\dz^+_{-\om}=1.
\nom}
From this system of equations, the coefficients $\xi^\pm_\om$ can be easily determined in terms of the already computed $\dz^\pm_\om$. Thus, we have established a method for numerically calculating the coefficients $\xi^\pm_\om$, which enables the numerical computation of the function $q_\om$ using formula \eqref{snh3}. We do not provide graphs of the coefficients $\xi^\pm_\om$ computed in this way, as they are not particularly informative due to their dependence on the specific definition of the confluent Heun function. However, the numerical values of the function $q_\om$ are independent of this choice.

\section*{Appendix~3}
The method for determining the coefficients $\dz^\pm_\om$ in formula \eqref{pra2}, described in Appendix~2, can also be applied to numerically calculate the coefficients $K^\pm_\om$ in the asymptotics \eqref{as8}, valid as $\chi\to-\infty$. Indeed, \eqref{as8} can be written in the form of \eqref{pra2}, if we substitute $K^\pm_\om$ for $\dz^\pm_\om$ and use the notation $u_\om=q_\om$ with
\disn{prb1}{
v^\pm_\om(r)=e^{\pm i\om\chi}\ls 1+\frac{e^{\frac{\chi}{R}-1}}{1\pm 2i\om R}\rs.
\nom}
Thus, analogous to \eqref{pra3}, we can write:
\disn{prb2}{
K^\pm_\om=\lim_{\chi\to-\infty}\ls\frac{\dd}{\dd r}\frac{q_\om}{v^\mp_\om(r)}\rs
\ls\frac{\dd}{\dd r}\frac{v^\pm_\om(r)}{v^\mp_\om(r)}\rs^{-1},
\nom}
where $v^\pm_\om(r)$ is defined by \eqref{prb1}.

\begin{figure}[htbp]\centering
\includegraphics[width=0.47\textwidth]{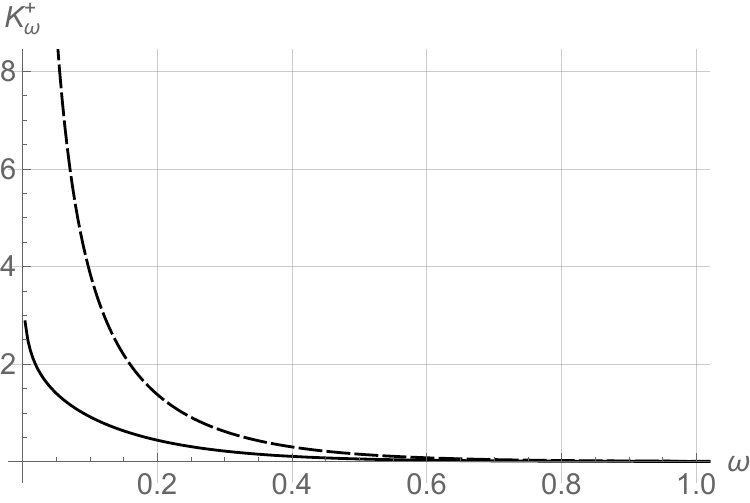}
\quad
\includegraphics[width=0.47\textwidth]{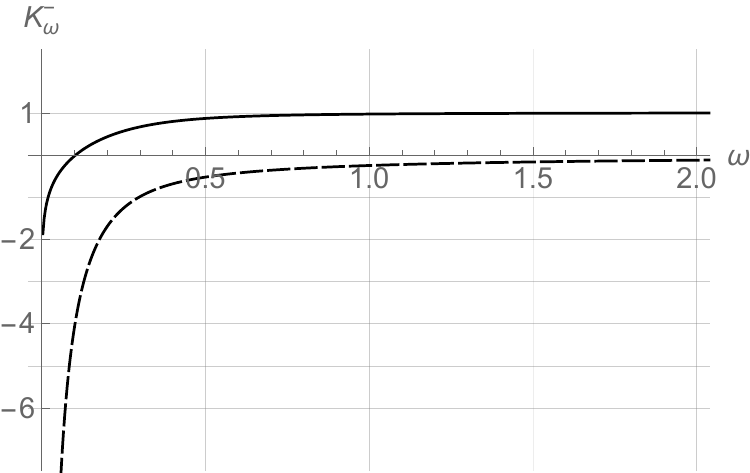}
\caption{Numerical results for the dependence of the coefficients $K^+_\om$ (left) and $K^-_\om$ (right)
on the frequency $\om$, plotted in units of $1/R$. Solid lines represent the real parts of the coefficients, and dashed lines represent their imaginary parts.}
\label{grafK}
\end{figure}

Numerical calculations show that the term under the limit in \eqref{prb2} converges to its limit very quickly, even significantly faster than in the case described in Appendix~2. This allows us to obtain $K^\pm_\om$ with good accuracy by evaluating the expression under the limit for a sufficiently large negative $\chi$. Moreover, it turns out that we can even forgo the refined asymptotics \eqref{as8} and use only the leading approximation, defining $v^\pm_\om(r)$ not by \eqref{prb1} but by its leading approximation:
\disn{prb3}{
v^\pm_\om(r)=e^{\pm i\om\chi}.
\nom}
The reason for this lies in the exponentially small nature of the correction in \eqref{prb1}. In our calculations, we used \eqref{prb3} and evaluated the term under the limit in \eqref{prb2} at $\chi=-24$. The graphs of the coefficients $K^\pm_\om$ obtained from numerical calculations are shown in Fig.~\ref{grafK}.

\section*{Appendix~4}
Here, we present the explicit forms of the functions \( f_{1,2,3}(r) \) used in the asymptotic expression for \( \bar{S}_\om \):
\disn{prf1}{
f_1(r)=\frac{1}{4s^3R}\ls\left(2 s+\sqrt{s (4 s+3)+1}\right) \sqrt{s \left(8 s-4 \sqrt{s (4 s+3)+1}+3\right)+1}\rs^{-1}\times \ns \times \left(
-8 s^3+s^2\left(-3+4\sqrt{1+s(3+4s)}\right)+2\sqrt{1+s\left(3+8s-4\sqrt{1+s(3+4s)}\right)}
+\right.\ns \left.+
s\biggl(4+3\sqrt{1+s(3+4s)}+\sqrt{1+s\left(3+8s-4\sqrt{1+s(3+4s)}\right)}-
\right. \ns \left. -3\sqrt{1+s(3+4s)}\sqrt{1+s\left(3+8s-4\sqrt{1+s(3+4s)}\right)}\biggr)
+\right. \ns \left. +\left(1+3\sqrt{1+s(3+4s)}\right) \left(-1+\sqrt{-1+\sqrt{1+s\left(3+8s-4\sqrt{1+s(3+4s)}\right)}}\right)
\right)
,
\nom}
\disn{prf2}{
f_2(r)=-\frac{1}{s} \left(-s+\frac{\sqrt{4 s^2+3 r+1}}{\sqrt{4 s^2+3 s+1}+2 s} -\frac{(s+1) \sqrt{8
   s^2+\left(3-4 \sqrt{4 s^2+3 s+1}\right) s+1}}{3 s+1} \right. +\ns \left.-\frac{s \left(2 \sqrt{4 s^2+3 s+1}-s-2\right)-1}{\sqrt{8 s^2+\left(3-4 \sqrt{4 s^2+3 s+1}\right) s+1}\left(\sqrt{4 s^2+3 s+1}+2 s\right)}-\frac{s+1}{\sqrt{4 s^2+3 s+1}+2 s}\right)
,
\nom}
\disn{prf3}{
f_3(r)=-\frac{1}{(s-1)R}\left( s-\frac{\sqrt{4 s^2+3 s+1}}{\sqrt{4 s^2+3 s+1}+2 s}-\frac{(s+1) \sqrt{8 s^2-4 \sqrt{4 s^2+3 s+1} s+3 s+1}}{3 s+1}-
\right. \ns \left. -\frac{s \left(2 \sqrt{4 s^2+3 s+1}-s-2\right)-1}{\sqrt{8 s^2+\left(3-4 \sqrt{4 s^2+3 s+1}\right) s+1}\left(\sqrt{4 s^2+3 s+1}+2 s\right)}+\frac{s+1}{\sqrt{4 s^2+3 s+1}+2 s}
   \right)
,
\nom}
where \( s\equiv r/R \).

The limits of these functions as \( r\to R \) are:
\disn{prf4}{
f_1(r)\str{r\to R}\frac{1}{R}\ls\sqrt{2}-1\rs,\qquad
f_2(r)\str{r\to R}4\ls\sqrt{2}-1\rs,\qquad
f_3(r)\str{r\to R}0,
\nom}
and their asymptotics as \( r\to\infty \) are:
\disn{prf5}{
f_1(r)\approx\frac{1}{4r^2},\qquad
f_2(r)\approx2-\frac{1}{2r},\qquad
f_3(r)\approx\frac{3}{4r^2}.
\nom}


\end{document}